# Esports Athletes and Players: a Comparative Study

**Nikita Khromov**

Skolkovo Institute of Science and Technology, Russia

**Alexander Korotin**

Skolkovo Institute of Science and Technology, Russia

**Andrey Lange**

Skolkovo Institute of Science and Technology, Russia

**Anton Stepanov**

Skolkovo Institute of Science and Technology, Russia

**Evgeny Burnaev**

Skolkovo Institute of Science and Technology, Russia

**Andrey Somov**

Skolkovo Institute of Science and Technology, Russia

We present a comparative study of the players' and professional players' (athletes') performance in Counter Strike: Global Offensive (CS:GO) discipline. Our study is based on ubiquitous sensing helping identify the biometric features significantly contributing to the classification of particular skills of the players. The research provides better understanding why the athletes demonstrate superior performance as compared to other players.

eSports is an organized and competitive gaming with a specific goal at the end of a game where single players or teams compete against each other. In spite of popularity and official recognition of esports, the debates as to the assessment of esports as an actual sport and comparing it to a sport are still going on[3]. The global esports audience numbered 380.2 million in 2018 and has tended to grow up to 557 million by 2021[(1)]. Apart from the rapid increase in the quantity of professional athletes and teams, the number of players has dramatically gone up: 27 million people play League of Legends every day[1]. In this research we define an *athlete* as a *professional* player with a work contract with a professional eSports team (we use *athlete* and *professional player* interchangeably throughout the article). A *player* is a person without the eSports contract while having relevant game skills or status.

In spite of popularity and official recognition of esports, the debates as to the assessment of esports as an actual sport are still going on. Although professional esports players spend 8-12 hours

---

[(1)] Newzoo. 2018 Global Esports Market Report. https://asociacionempresarialesports.es/wp-content/uploads/newzoo_2018_global_esports_market_report_excerpt.pdf

a day for their training, video games are nevertheless considered in the society as a sort of entertainment. At the same time, esports research is in its infancy - we still do not know how to efficiently conduct trainings of esports athletes[14] and how to compose the teams - the coaches rely on their professional experience rather than on scientific approach.

In fact, a number of online services[(2)] are able to provide the generic in-game statistics, e.g. the rank evolution, headshots percentage and match win average, for further analysis. However, there are no tools available for the detailed physiological and in-game analysis to prove the qualification of a player. The recent advances in pervasive sensing and computing that shape the emergence of pervasive data science[2], as well as the involving of professional players into the esports research, will help identify the factors for defining the specific skill level of each particular player.

The present research is done in collaboration with the professional esports team Monolith[(3)], Russia, in the scope of Skoltech Cyberacademy activity. Our collaboration ensures practical feasibility of this work and allows researchers to dig deeper into the details of esports. In this article we perform a multivariate analysis of data collected from an eye tracker, keyboard and mouse while subjects (athletes, players and newbies) play CS:GO. We then investigate features contributing to the classification of those subjects according to their gaming skills.

## RELATED WORK

Existing research efforts in esports lack the experimentation and the involvement of professional players. Although the research which is going through its infancy has been conducted so far without taking into account the multidisciplinary approach, we summarize some recent advances in the area.

The lion share of the current research in esports is carried out in the scope of affective computing in games. It is the interdisciplinary field where the community works on modelling and development of systems able to recognize, process and simulate human affect. The detailed discussion on the theoretical reasons to favor ordinal labels for the annotation and representation of emotions is provided by Yannakakis et al.[4]

Research on prediction of the player skill upgrading within a gaming season[5] and forecasting the player behavioral data[6] is reported recently. Aung et al. investigate how the player performance depends on skill learning[5]. For this analysis the authors build two multivariate classifiers. At the end of the research the authors come to the conclusion that the final performance in a game under investigation has a strong relationship with early skill learning. Guitart et al. perform an experimental analysis using artificial intelligence methods for daily forecasting of playtime and sales[6]. The research outcome is as follows: deep learning[10] could be used as a general model for forecasting various time series characterized by the different dynamic behavior.

Another important research direction is associated with the investigation of social structures in player groups [7]. This work presents the analysis of correlation with the aim to identify the effect of players' group characteristics on group activity. For conducting the analysis the authors combine the information from the social network with self-report information available at a social matchmaking service across the players of the online first person shooter Destiny. This paper is characterized by integrating demographic and preference data apart from considering the information from a player established community only.

Almeida et al. carried out a study[11] where they applied the eye-tracking system for understanding how the players visually interact with a game scenario. Different groups of players were compared and differences between the experienced and inexperienced players were shown. If these differences are known for a specific game, there is an opportunity to rely on this feature in the experimental setup and use it for recruiting of players.

However, the gameplay input recorded using a computer mouse and the buttons of keyboard remains the main source of collecting data on and modelling the players behavior and their ability prediction[8]. The appearance of wearables and 'earables'[9] makes them an excellent candidate

---

[(2)] CS:GO Demos Manager, https://csgo-demos-manager.com
[(3)] Monolith team profile, https://www.hltv.org/team/9182/monolith

for upgrading the behaviour models. Indeed, the running machine learning algorithms on the resource constrained devices[10], e.g. wearables, is a promising research direction for many domains including esports and game analytics[13].

## ESPORTS BACKGROUND

The situation with gaming has changed dramatically over the last decade. According to recent forecasts 60 percent of Americans play video games daily, 41 percent of players use personal computers and first-person shooters are the most popular multiplayer games.

In most eSports disciplines, an instant assessment of the game situation, the fastest possible reaction over several game rounds and concentration on the game for a long time are essential for the professional esports players. A CS:GO professional team typically includes 5 athletes and a coach. The coach objectives are to help the athletes in analyzing their game, devising the team strategy and tactic, assigning the in-game roles, to cheer the team, to develop the players' skills and to scout for new team members.

Apart from the skill metrics used in the teams, there is the CS:GO ranking system for each single player. There are currently 18 ranks[(4)] in CS:GO starting from the lowest "Silver 1" to the highest "The Global Elite". The "1-18" rank scale system was used in this research for the better data interpretation.

## DATA COLLECTION

The goal of our experiment is to collect biometric data from athletes and players and to perform a comparative study of the data obtained. This includes engineering of relevant features and identifying the feature importance using methods of statistics and machine learning.

We created a **testbed** (see Figure 1) which is the complete gaming place with integrated sensors for recording the biometric data.

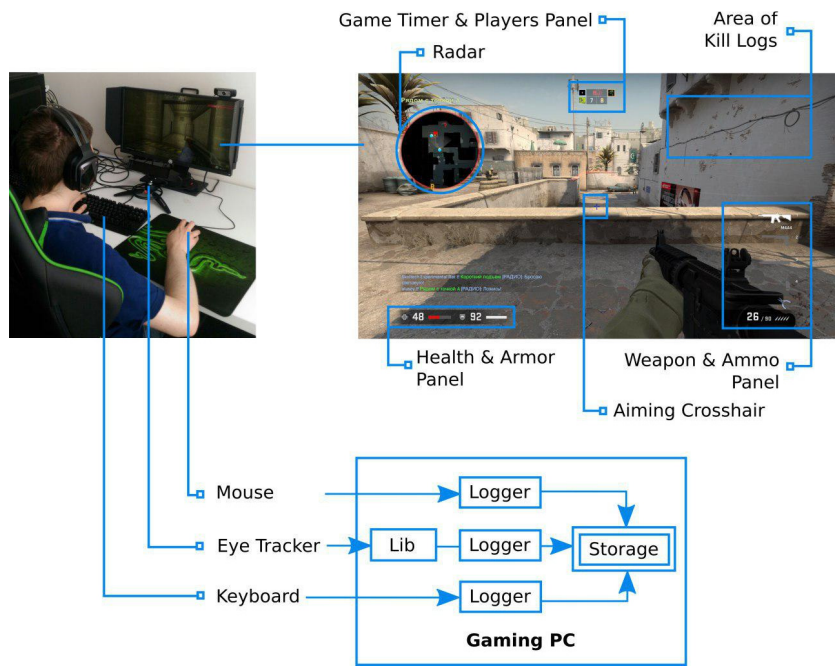

Figure 1. Experimental testbed, its block diagram and a print screen of CS:GO game scene.

---

[(4)] Ranking system in CS:GO, https://csgo-stats.com/ranks

The experimental testbed includes several sensors: the eye tracker EyeLink, the mouse and the keyboard loggers (see Figure 1). The data from different sources is synchronized to the desired tick rate and is interpolated.

**The eye tracker** relies on the SR Research library which performs the analysis of the position of each eye and calculates the average coordinates relative to the edge of the screen. The EyeLink tracker provided the eye position, the gaze point data and time stamps. It has 250 Hz sampling rate. It was calibrated before every game session for each player for high measurement accuracy.

**The keyboard and mouse** data is recorded during every game session with the 10 ms period by our custom software to prevent any conflicting situations with running game and to avoid the system overload. The mouse coordinates, as well as the pressed mouse and keyboard buttons, are recorded.

## Game Scenario

CS:GO is an objective-based, multiplayer first-person shooter. In game, two opposing teams - the Terrorists and the Counter Terrorists - compete in different game modes to complete objectives, such as securing a location to plant or defuse a bomb and rescuing or guarding hostages. Every play control one member of one of the teams (counter-terrorists or terrorists). An example of player's in-game view and user-interface elements is shown in Figure 1.

The main competitive regime used in tournaments is bomb plant/defuse. The other popular game regimes include deathmatch, retake, surfing, zombie mod among others. They provide the players with many tactical and strategic decisions to make during each game: location to go, a weapon to buy, role to play, etc.

For our study, we chose the Deathmatch (DM) modification. In DM, the goal of the player is to achieve as many kills of other players as possible keeping the number of own deaths as low as possible. If the model of the player is killed, it immediately spawns again according to DM rules. The set of potential weapons to use is usually defined by the game server and remains the same for each player on the server.

There are several reasons why we chose DM modification. First of all, this modification is used by professional eSports players regularly as part of their training process. It helps players to improve reaction and aiming skill. Secondly, while playing DM, the player is placed in the same situation many times. It removes the necessity of taking into account the impact of other factors, such as team play (there are no teams in DM), radar controlling (the radar is disabled), game economics (the equipment is given automatically), etc.

## Participants

We invited 28 subjects to participate in our experiment (4 athletes and 24 players). Every participant signed a consent form which allowed recording data from the game and physiological data from sensors.

In our experiment, the professional class was represented by 4 players from **Monolith** professional team ranked from 65 to 100 in the world ranking according to HLTV[5] resource. All of them are males from 19 to 25 years old. An average athlete from Monolith eSports has spent 7 years playing the CS:GO at least 8 hours per day and achieved the highest in-game rank.

Besides eSports athletes, 24 players (21 males, 3 females) were recruited through the online advertising at the institute internal website or posters to take part in the experiment. These players were also questioned about their previous gaming experience in CS:GO and split into 3 subgroups:

- High-skill amateurs: 11 people having more than 700 hours. They show high performance and try hard, but without focusing on the eSports career;

---

[5] HLTV resource, https://www.hltv.org.

- Low-skill amateurs: 7 people having from 10 to 700 gaming hours. They play for fun and have some gaming experience;
- Newbies: 6 people having less than 10 gaming hours, i.e. without much gaming experience.

## DATA ANALYSIS

In this section we describe the features that characterize every period of the game for each player and the model for predicting the level of player proficiency.

Each of 28 participants played the game session in CS:GO of about 30 minutes. The biometric-based features described next were calculated for every rolling time window of 5 minutes width taken after every 30 seconds step. It allowed us to construct the dataset of 900 valid samples (161 samples from professional players), where each feature vector corresponds to the time window of the particular player game session.

In this work, we perform the binary classification considering 2 classes of qualification: athletes vs. players. The reasons are as follows: (i) the actual problem is to understand how a player has become closer to the athletes, (ii) only professional players class labeling is proved in our experiment.

### Eye-tracking Based Features

Since CS:GO is a first-person shooter the players mainly look at the center of the screen where the aiming crosshair is located. Nevertheless, there are other areas of interest including user interface (UI) elements (see Figure 1). The Deathmatch game mode minimizes the necessity of looking at the UI elements because the radar is disabled and the health, armor, and ammo are restored automatically after every kill or death. Therefore, the player's gaze is mainly concentrated on the game environment (objects, enemies, shooting, walking) and aiming crosshair rather than on the UI elements.

Our experiment shows that there are significant differences in the player's gaze between the skill groups. It turns out that the higher skill the player has, the more the gaze is concentrated on the screen center. To illustrate this, we build the heatmaps for the players of different skill (see Figure 2). The heatmaps are obtained from 2D histograms with additional smoothing using Gaussian kernel density estimation.

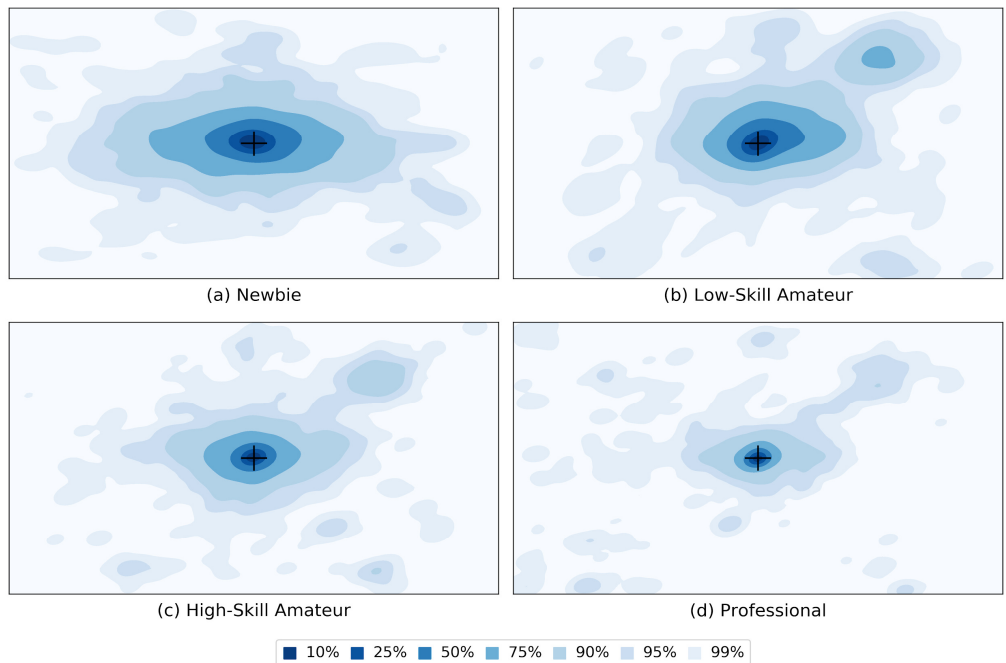

Figure 2. Gaze heatmaps for players of different skill. Every color area bounds the cumulative probability of the gaze to be inside it.

There are two main reasons why high skilled players spend more time looking into the center of the screen. First of all, they have better knowledge of the game map and always know how to position themselves and where to aim. Therefore, they do not look around the screen frequently. Secondly, in the situation when an enemy appears (not in the expected screen center position), the skilled players much more quickly move the in-game crosshair to the enemy and then again look at the crosshair keeping the enemy on sight.

Thus, to characterize the player's gaze during the time window we compute the mean deviation of player's gaze from the center of the screen.

## Keyboard and Mouse Based Features

We describe the keyboard and mouse buttons data streams by different features of two types:

- The percentage of the time when a combination of keys and mouse buttons was pressed (the total time when the combination was pressed divided by the width of the full time of a game session or a time window), it is between 0 and 1. It characterizes the use (or "usage") of specific control elements and does not distinguish if a combination was used frequently keeping pushed for a short time, or rarely and keeping pushed for a long time;
- The average length of continuous time interval (in seconds) when the specific keys and mouse buttons were pressed.

The main controls in game are the keys: **W** (forward), **S** (back), **A** (left), **D** (right), **Ctrl** ("duck", i.e. squat pose) and the **MOUSE1** – left mouse button (weapon fire). Specific key/button combinations imply additional logic: "A & Ctrl & MOUSE1" means that at least these 3 controls are pressed together, "A or D" ("W or S") means that A or D (W or S) are among the controls pressed at that moment.

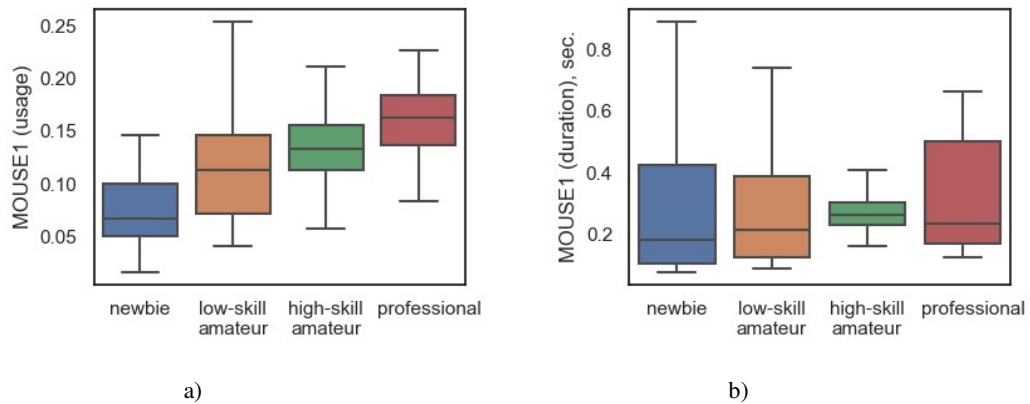

a)                    b)

Figure 3. Distributions of (a) the mouse usage (the usage of weapon fire) and (b) the duration of weapon fire, for different groups of subjects.

Some features relevant to a player's skill level are shown in Figure 3. Figure 3a shows the *usage* of the left mouse button. The athletes fire more actively, which is expected. Figure 3b shows the *duration* of left mouse button keeping pressed. Unlike to the *usage*, the *duration* does not differ significantly among the players' groups.

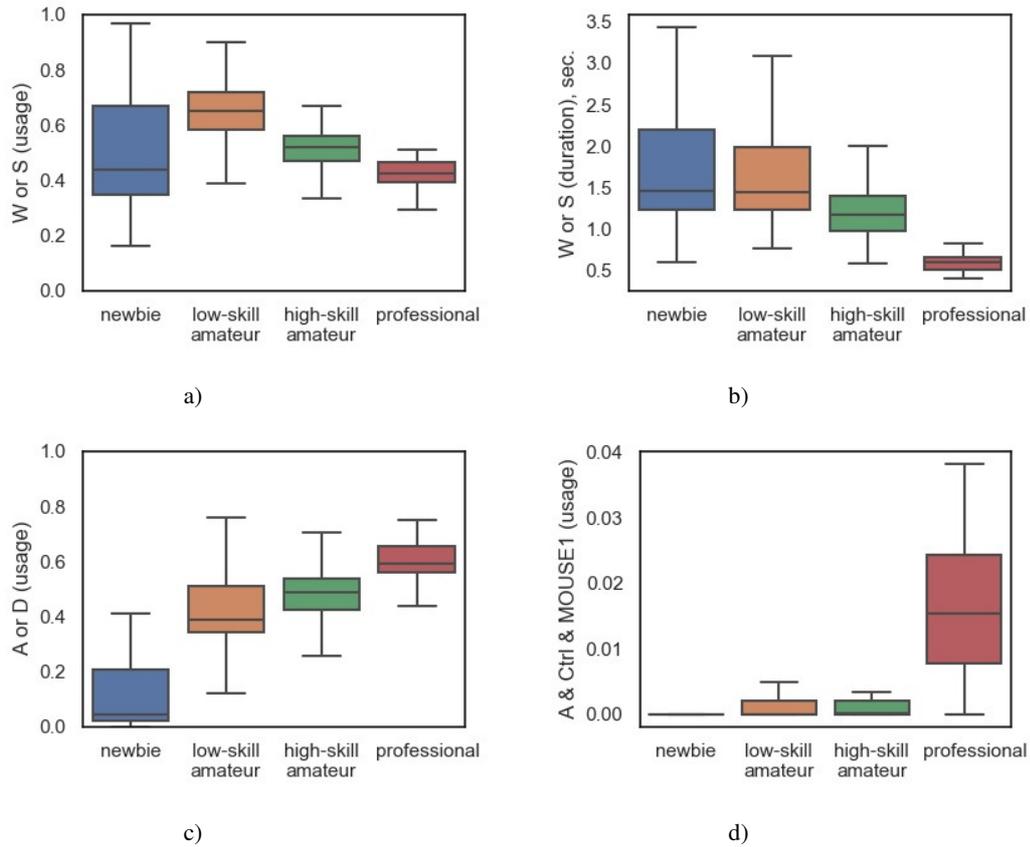

Figure 4. Distributions of the most important key/mouse features for different groups of subjects. (a) The usage of forward and backward motions, (b) the duration of forward and backward motions, (c) the usage of left and right motions, (d) the usage of the "shoot being in the duck pose moving to the left" technique.

Here are some more interesting insights into behavior of different players when using control keys. Quite unexpectedly, professional players use forward, and backward motions on average like newbies (see Figure 4a) and the usage of these motions decreases as the experience of the subject grows from the low-skill amateur to the professional. The duration of these motions decreases as the subject skill increases (see Figure 4b). Moreover, since the class of professional players appears quite compact and separable from other sub-classes, it means that this feature can be the most important for the player skill prediction.

Along with the fact that professional players have less usage of forward and backward motions, they move left and right more often than the players from other groups (see Figure 4c). Moreover, the higher this feature, the higher the skill.

Also, the usage and duration of more exclusive combinations peculiar to professional players, such as "A & Ctrl & MOUSE1", are considered (see Figure 4d). This feature is comparatively exclusive, i.e., the newbies almost do not know about it and only a few amateurs exploit it. Even some professional players use it occasionally, which results in high variance on the box-plot. Nevertheless, the high value of this feature indicates the highly skilled player.

In addition, we consider the usage of any single key "1keys (usage)" as well as the duration of the left and right motions "A or D (duration)". Without showing their distributions here it is worth noting that the professional players have a slightly higher mean of "1keys (usage)" (0.5 vs. 0.42) and the relationship of "A or D (duration)" with the skill level is quite poor. Indeed, the distributions of A and D durations are quite similar. The difference is that the newbies move left and right only accidentally while the professional players' key pressings are usually quicker.

## Statistical Significance of Features

In order to justify the visual insights from the feature distributions shown in previous section we performed the statistical tests. For every feature we checked the null hypothesis that two distributions are the same with the alternative being that one distribution is stochastically greater than the other. Two samples are independent since they are from different classes of players. We aim at making the values independent within each sample when we increased the rolling window step from 30 seconds to 5 minutes to make windows (of 5 minutes width) non overlapping and placed exactly one after one. Given that both samples are reduced and have different variances, as well as without assumption of normality, the Mann–Whitney test[15] was used. The p-values of the features are shown in Table 1.

Table 1. Mann-Whitney test for athletes vs. players (significant features are highlighted).

|  | 1keys [usage] | Mouse1 [usage] | Mouse1 [duration] | W or S [usage] | W or S [duration] | A or D [usage] | A or D [duration] | A & Ctrl & Mouse1 [usage] | Gaze [std] |
|---|---|---|---|---|---|---|---|---|---|
| p-value | <0.001 | <0.001 | 0.357 | <0.001 | <0.001 | <0.001 | 0.025 | <0.001 | <0.001 |

According to the test results, all features except for the "MOUSE1 (duration)" and "A or D (duration)" are the statistically significant factors with the significance level 0.01. It is quite expected: the important features, i.e. features demonstrating the visual relationship between their values and the class of the skill in Figure 3 and Figure 4 tend to have smaller p-values.

## Predicting the Player Skill

The model for predicting the player's skill was developed using the set of biometric-based features. This model provides more confidence evaluation of the features' relevance to the subject proficiency than the visual analysis of distributions and the statistical testing discussed before. Unlike them, it evaluates the whole set of features together. Moreover, the probability that the given feature vector belongs to the class of professional players allows us to estimate the overall "biometric skill" of the subject including the dynamics of key pressings, etc.

The dataset with 900 samples and 9 features was used for training. We used Extremely Randomized Trees classifier[12] that builds an ensemble of totally randomized decision trees. Its robustness with respect to the class labels is crucial when the dataset is rather small. Class balancing is provided by sampling. The hyperparameters include the number of trees (from 10 to 1000), the maximal depth (from 1 to 8) and the bootstrap (either used or not). The overfitting is controlled through the leave-one-player-out cross-validation procedure.

The best overall validation accuracy 0.9 is reached when the number of trees is 500, maximal tree depth is 1 (decision stump) and the bootstrap is used. Other performance metrics and the confusion matrix are shown in Figures 5a and Figure 5b, respectively.

Also, our classifier (as any decision tree model) provides the importance scores which show the real predictive power of every feature (see Figure 5c). The score of the predictor estimates the contribution of its splits during every decision tree construction to the model quality. The features are not highly correlated: the top absolute correlations are 0.59 ("1keys (usage)" and "Gaze (std)") and 0.56 ("W or S (usage)" "W & S (duration)"). It is worth noting that the maximum value corresponds to the features of even different nature (gaze vs. keys), although the mouse and keys pairs of features have smaller values.

|  | Precision | Recall | F1 score | ROC AUC | Accuracy |
|---|---|---|---|---|---|
| non-pro | 0.96 | 0.92 | 0.94 | 0.97 | 0.9 |
| pro | 0.69 | 0.82 | 0.75 | | |

(a)

|  | Predicted non-pro | Predicted pro | Support |
|---|---|---|---|
| non-pro | 680 | 59 | 739 |
| pro | 29 | 132 | 161 |

(b)

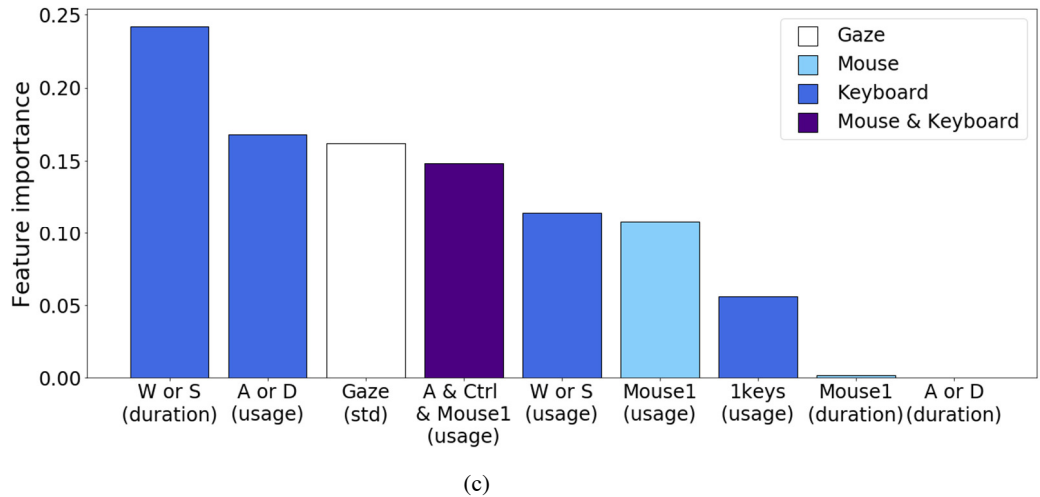

(c)

Figure 5. Results of the classification and the importance of the features: (a) performance metrics, (b) confusion matrix, (c) feature importance.

As we can see in Figure 5a, the order of features is in line with the visual analysis of distributions shown in Figure 3 and Figure 4 and with the statistical test shown in Table 1: the better the feature separates athletes from players the higher predictive importance it has.

## CONCLUSIONS

Data science and machine learning have dramatically progressed over the last decade and provide great support in designing our physical environment. They help with problem solving in numerous computing applications, in particular the cutting-edge applications such as esports.

In this work, we designed the biometric features helping distinguish the professional CS:GO athletes from the non-professional players as well as to estimate their classification power using statistical methods. The bottom line of this research is to identify the specific in-game behavior and to measure the individual level of the athlete proficiency from the physiological perspective. The results of this work let the professional esports teams to rely on science in addition to their experience.

## ACKNOWLEDGMENTS

The reported study was funded by RFBR according to the research project No. 18-29-22077.

Authors would like to thank Skoltech Cyberacademy, CS:GO **Monolith** team and their coach Rustam "**Tsa-Ga**" Tsagolov for fruitful discussions while preparing this article. Also, the authors thank Alexey **"ub1que"** Polivanov for supporting the experiments by providing a slot at the CS:GO Online Deathmatch server (http://ub1que.ru).

We would like to thank the reviewers for a thoughtful analysis of our manuscript and for the useful comments helped to improve the article.

## ABOUT THE AUTHORS

**Nikita Khromov** is a PhD student at Skolkovo Institute of Science and Technology (Skoltech), Russia. His research interests include computer vision, machine learning and human-computer interaction. Nikita received his MS degree in Fundamental Informatics and Information Technologies from the Peoples' Friendship University of Russia (PFUR). Contact him at nikita.khromov@skoltech.ru.

**Alexander Korotin** is a PhD student Skoltech, Russia. Contact him at a.korotin@skoltech.ru.

**Andrey Lange** is a Senior Research Engineer at Skoltech, Russia. His research interests include data science modeling in different applications. Dr. Lange received his PhD in Proba-bility Theory and Statistics from Bauman Moscow State Technical University, Russia. Contact him at a.lange@skoltech.ru.

**Anton Stepanov** is a Research-Engineer at Skoltech, Russia. His research interests include data acquisition systems and experiments automation. Contact him at a.stepanov@skoltech.ru.

**Evgeny Burnaev** is an Associate Professor at Skolkovo Institute of Science and Technology (Skoltech), Russia. Contact him at e.burnaev@skoltech.ru.

**Andrey Somov** is an Assistant Professor at Skolkovo Institute of Science and Technology (Skoltech), Russia. His research interests include wearable sensing and cognitive Internet of Things. Dr. Somov received his PhD in Electronic Engineering from the University of Trento, Italy. Contact him at a.somov@skoltech.ru.